\pdfoutput=1
\documentclass[11pt]{article}

\usepackage[preprint]{acl}

\usepackage{times}
\usepackage{latexsym}
\usepackage{multicol}
\usepackage{multirow}
\usepackage{tabularx}
\usepackage{booktabs}
\usepackage{subcaption}
\usepackage{float}
\usepackage{balance}
\usepackage[T1]{fontenc}
\usepackage[utf8]{inputenc}

\usepackage{microtype}

\usepackage{inconsolata}

\usepackage{graphicx}

\title{S2SBench: A Benchmark for Quantifying Intelligence Degradation in Speech-to-Speech Large Language Models}

\author{
 \textbf{Yuanbo Fang\textsuperscript{1,2}\thanks{This work was done during an internship at Baichuan Inc.}},
 \textbf{Haoze Sun\textsuperscript{2}},
 \textbf{Jun Liu\textsuperscript{2}},
 \textbf{Tao Zhang\textsuperscript{2}},
\\
 \textbf{Zenan Zhou\textsuperscript{2}},
 \textbf{Weipeng Chen\textsuperscript{2}},
 \textbf{Xiaofen Xing\textsuperscript{1}\thanks{Corresponding author.}},
 \textbf{Xiangmin Xu \textsuperscript{1}},
\\
 \textsuperscript{1}South China University of Technology,
 \textsuperscript{2}Baichuan Inc.
}

\begin{document}
\maketitle
\begin{abstract}
End-to-end speech large language models ((LLMs)) extend the capabilities of text-based models to directly process and generate audio tokens. However, this often leads to a decline in reasoning and generation performance compared to text input, a phenomenon referred to as intelligence degradation. To systematically evaluate this gap, we propose S2SBench, a benchmark designed to quantify performance degradation in Speech LLMs. It includes diagnostic datasets targeting sentence continuation and commonsense reasoning under audio input. We further introduce a pairwise evaluation protocol based on perplexity differences between plausible and implausible samples to measure degradation relative to text input. We apply S2SBench to analyze the training process of Baichuan-Audio, which further demonstrates the benchmark's effectiveness. All datasets and evaluation code are available at \url{https://github.com/undobug/S2SBench}.

% This document is a supplement to the general instructions for *ACL authors. It contains instructions for using the \LaTeX{} style files for ACL conferences.
% The document itself conforms to its own specifications, and is therefore an example of what your manuscript should look like.
% These instructions should be used both for papers submitted for review and for final versions of accepted papers.
\end{abstract}

\section{Introduction}
LLMs have achieved remarkable success across diverse natural language processing tasks and have recently expanded to multimodal domains, including vision-language \cite{lin2023video,li2024baichuan}and audio-language\cite{hsu2021hubert,chen2022wavlm,chu2024qwen2,hassid2024textually}. Models like Qwen\cite{qwen2025qwen25technicalreport} and the GPT \cite{achiam2023gpt} series demonstrate strong generalization, in-context learning, and reasoning abilities. Recently, there is growing interest in enabling LLMs to directly process and generate speech, spurring development of end-to-end Speech Large Language Models\cite{HelloGPT4o,li2025baichuan,fang2024llama,yao2024minicpm,fu2025vita,yu2024salmonn}. 

In contrast to the conventional cascaded pipeline comprising ASR, LLM, and TTS modules, end-to-end Speech LLMs are designed to process raw audio inputs and generate outputs directly, without relying on intermediate transcription or synthesis\cite{hassid2024textually,nguyen2025spirit}. This unified architecture simplifies the processing flow and preserves prosodic and speaker-specific cues. While most existing models leverage pretrained text-based LLMs and inherit their strong language understanding capabilities, recent studies report a notable performance gap when these models are applied to spoken language tasks. This degradation, often referred to as intelligence degradation, is primarily attributed to three factors\cite{wang2024speech}: the limited semantic density of audio tokens, longer sequence lengths relative to text, and variability introduced by prosody and speaker characteristics. These challenges hinder the model’s ability to construct coherent internal representations, thereby reducing overall performance.

\begin{figure*}[t]
  \centering
  \includegraphics[width=16cm]{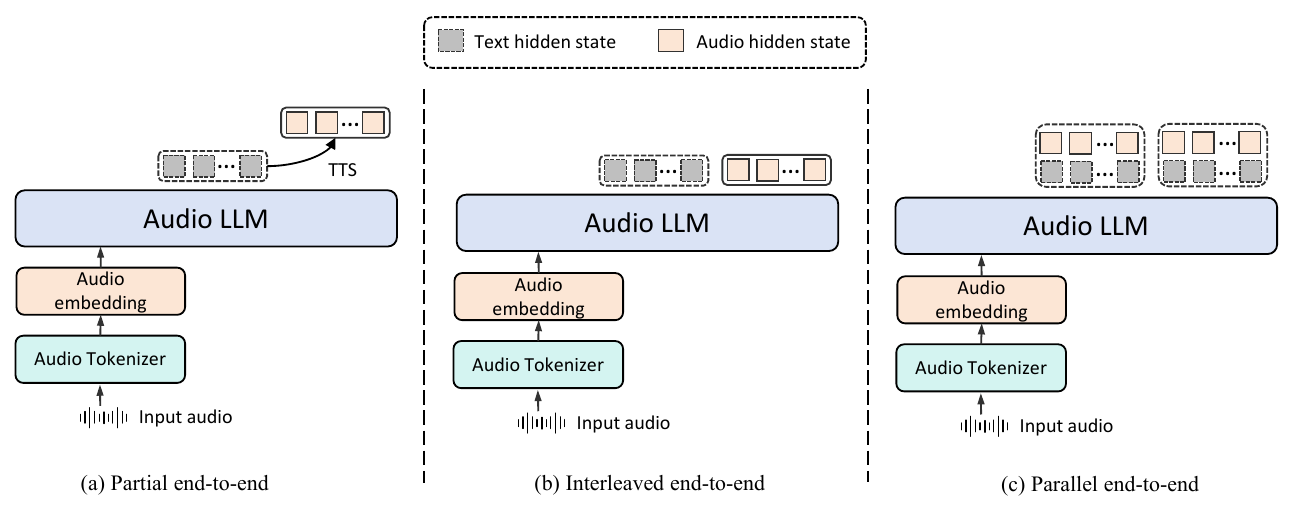}
  \caption{Architectural types of end-to-end Speech LLMs: (a) Partial end-to-end, (b) Interleaved fully end-to-end, and (c) Parallel fully end-to-end.}
  \label{fig:3model}
\end{figure*}

To systematically investigate the extent and nature of intelligence degradation in end-to-end Speech LLMs, we introduce S2SBench, a benchmark designed to quantify performance loss under audio input. Taking performance under text input as a reference, S2SBench evaluates model degradation across three diagnostic datasets focused on sentence continuation and commonsense reasoning. To support fine-grained analysis, we further propose a pairwise evaluation method based on perplexity differences between semantically plausible and implausible samples. Our contributions are as follows:

(1) We introduce a benchmark framework for quantifying intelligence degradation in end-to-end Speech LLMs by comparing performance under audio and text input.

(2) We construct evaluation datasets that focus on sentence continuation and commonsense reasoning to evaluate the intelligence capability of LLMs.

(3) We validate the effectiveness of the proposed S2SBench framework by applying it to evaluate the two-stage training strategy used in Baichuan-Audio.

\section{Background}

End-to-end Speech Large Language Models (E2E Speech LLMs) aim to extend language models to directly process and generate speech, bypassing intermediate transcription and synthesis stages. Existing approaches can be grouped into three categories, as illustrated in Figure~\ref{fig:3model}: partial end-to-end, interleaved fully end-to-end, and parallel fully end-to-end.

\textbf{Partial end-to-end} models, such as Freeze-Omni~\cite{wang2024freeze}, integrate pretrained text-based LLMs with modality adapters and speech decoders. Audio input is aligned with the text modality through an audio encoder, enabling the language model to operate on hidden states that resemble text tokens. Speech output is generated by decoding hidden representations into waveforms. This design preserves the language modeling capabilities of the LLM but lacks unified modeling of the audio modality and does not support full end-to-end understanding.

\textbf{Interleaved fully end-to-end} models, including Baichuan-Audio\cite{li2025baichuan} and GLM-4-Voice~\cite{zeng2024glm}, take a different approach. These models generate alternating audio and text tokens, allowing text segments to guide the speech generation process. This improves coherence and output quality while maintaining a tighter integration between language and audio processing.

\textbf{Parallel fully end-to-end} models, such as Moshi~\cite{defossez2024moshi}, directly operate on discrete audio tokens. Using a multi-stream architecture, these models simultaneously generate both speech and text outputs from audio inputs. They remove explicit textual grounding entirely, relying solely on audio-token representations for both understanding and generation.

Current mainstream end-to-end speech large model paradigms mostly build upon pre-trained pure-text large language models (LLMs), incorporating the audio modality to enable speech understanding and generation capabilities. However, fully end-to-end models enable more natural interaction but often suffer from intelligence degradation due to the structural gap between audio and text tokens. A systematic evaluation of intelligence degradation in these models remains missing.

\section{Evaluation Framework}

To systematically assess the intelligence degradation introduced by audio token inputs in end-to-end Speech LLMs, we propose an evaluation framework that contrasts model behavior across text-token and audio-token conditions. This section describes in detail the construction of the evaluation set and the evaluation method.

\subsection{Construction of Evaluation Datasets}

To comprehensively assess the intelligence capabilities of LLMs under different input modalities, we construct two types of evaluation datasets corresponding to two core tasks: Sentence continuation and commonsense reasoning. Each sample contains both text and audio modality versions, supporting comparative analysis of model performance across input formats. 

\textbf{Sentence continuation.}
For the continuation ability evaluation, we use the sStoryCloze dataset \cite{hassid2024textually}. Additionally, we introduce the zh-sStoryCloze dataset, which is created by translating the English version of sStoryCloze into its Chinese counterpart via a translation engine and replacing English names with Chinese ones to better suit the Chinese context. Each sample in both evaluation sets consists of five sentences, divided into positive and negative samples. The last sentence differs between the two, with the last sentence of the positive sample being the correct continuation. A prediction is considered correct if the perplexity of the last sentence in the positive sample is lower than that of the negative sample.

\textbf{Commonsense reasoning.}
For the commonsense reasoning ability evaluation, the goal is to assess whether the model possesses domain-specific knowledge. Drawing inspiration from the design of sStoryCloze, we use the GPT-4o API to rewrite and filter the CMMLU dataset \cite{li2023cmmlu}, ultimately creating the sCMMLU dataset with 4,743 commonsense questions. For each multiple-choice question in the original CMMLU, we rewrite it into four statements with the same first half and different second halves according to the answer options. A prediction is considered correct if the perplexity of the correct option's statement is lower than that of the other options.

\subsection{Evaluation method}

\begin{figure}[htb]
  \centering
  \includegraphics[width=7cm]{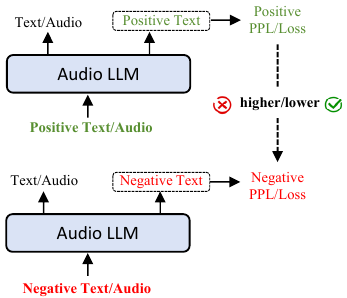}
  \caption{Evaluation pipeline for assessing the intelligence capability of large language models. The model architecture and reasoning task are identical under both text and audio input conditions.}
  \label{fig:ppl}
\end{figure}

To quantify the performance degradation introduced by the inclusion of the speech modality, comparative experiments are conducted under controlled settings to assess the impact of input modality on reasoning performance. The model structure and inference process remain unchanged in both configurations to ensure consistency. In the text-based setting, the model receives raw textual input, which is tokenized into text tokens and subsequently processed by the model. In the audio-based setting, the raw audio input is first discretized into audio tokens using the audio tokenizer module, and the resulting tokens are then passed into the model. An overview of the evaluation pipeline is presented in Figure~\ref{fig:ppl}.

Each evaluation instance contains a pair of samples, consisting of one positive example and one adversarial negative example. The positive example preserves logical, temporal, or commonsense consistency, whereas the negative example is deliberately constructed to disrupt semantic coherence or commonsense plausibility. Perplexity (PPL), a standard metric in language modeling, is employed to independently evaluate the plausibility of each sample. Lower perplexity indicates higher model confidence in a given sample. The model is considered to make a correct judgment when the positive example receives a lower perplexity than the corresponding negative example. Accuracy is computed as the proportion of instances in which the positive sample is assigned a lower perplexity than the negative sample. In addition to accuracy, the gap in perplexity between positive and negative examples is also examined to provide further insights into the model capability. A larger gap reflects a stronger distinction in model confidence and indicates enhanced reasoning performance.

\begin{table*}[t]
    \caption{\textbf{Performance Comparison on Various Evaluation Tasks.} $\ast$: Evaluations were performed using the instruct model as no base model was provided.}
    \label{tab:iq-performance}
    \centering
    \begin{tabular}{@{}lccccc@{}}
        \toprule
        \multirow{2}{*}{\textbf{Model}}   & \multirow{2}{*}{\textbf{Modality}} & \multirow{2}{*}{\textbf{Params}} & \multicolumn{3}{c}{\textbf{Evaluation Datasets}} \\ \cmidrule(l){4-6}
                         &                   &                 & \textbf{sStoryCloze} & \textbf{zh-sStoryCloze} & \textbf{sCMMLU} \\ \midrule
        TWIST            & S $\to$ T         & 7B               & 53.3                  & -                       & -              \\
        Moshi            & S $\to$ T         & 7B               & 60.8                  & -                       & -              \\
        GLM-4-Voice      & S $\to$ T         & 9B               & 76.3                  & 70.3$^{\ast}$                       & 64.3$^{\ast}$              \\
        \midrule
        Qwen2.5         & T $\to$ T         & 7B               & 83.0                  & 76.1                    & 70.3           \\
        Baichuan-Audio (single stage)  & S $\to$ T         & 7B               & 77.5                  & 70.1                    & 67.0           \\
        Baichuan-Audio (two stage)     & S $\to$ T         & 7B               & 79.6                  & 72.4                    & 69.3           \\ \bottomrule
    \end{tabular}
\end{table*}

\section{Experiments}

\subsection{Experimental Setup}

To evaluate the change in intelligence capability of end-to-end speech large language models during training, we conduct experiments on three benchmark datasets: English story continuation, Chinese story continuation, and commonsense reasoning. Each sample includes either a speech or text input, along with two candidate textual completions: one positive and one negative. The model is asked to identify the more appropriate option, which reflects its intelligence performance at that training stage.

We adopt two evaluation settings: speech-to-text and text-to-text. The semantic content of both modalities is kept consistent, and the evaluation tasks and metric computations are identical. We evaluate the models at multiple training checkpoints and report three metrics: accuracy, positive sample loss, and negative sample loss, which together reflect the development of the model's reasoning capability.

\subsection{Model Configuration}

Our experiments are based on the Baichuan-Audio model. We focus on the evolution of reasoning ability in the language model component during pretraining. The total number of training tokens is 109B. We compare two training strategies:
\begin{itemize}
  \item \textbf{Two-stage training}: In the first stage, the language model parameters are frozen, and only the audio embedding layer and audio head are updated. In the second stage, all parameters except the LM embedding layer and LM head are unfrozen for joint training.
  \item \textbf{Single-stage training}: All parameters are updated jointly from the beginning of training.
\end{itemize}
This comparison allows us to examine how different training strategies affect the development of the model's reasoning ability.

\subsection{Results}

The overall intelligence evaluation results for both settings are presented in~\autoref{tab:iq-performance}. As Baichuan-Audio is built upon Qwen2.5, its intelligence capability in the T$\to$T setting is theoretically constrained by the performance of the underlying text-only LLM. In contrast, the S$\to$T setting consistently shows lower accuracy due to the inherent challenges associated with processing audio tokens. Our benchmark specifically targets fully end-to-end speech models, as illustrated in Figure~\ref{fig:3model}. We do not include partial end-to-end models in our evaluation, as they primarily function in the semantic space and preserve the intelligence capability of their text-based backbone without notable degradation.

Baichuan-Audio aims to reduce the performance gap between S$\to$T and T$\to$T by improving the former toward the upper bound defined by the latter. We observe that the two-stage training strategy effectively mitigates intelligence degradation when compared to single-stage training. In the S$\to$T setting, the two-stage approach results in more stable training, clearer separation of positive and negative samples, and consistently higher accuracy. This strategy reduces interference caused by audio tokens and helps preserve the pretrained knowledge encoded in the language model.

In addition, we conduct step-wise evaluation during training to visualize changes in intelligence capability over time. As shown in the appendix, this dynamic evaluation provides further insight into model convergence and supports the identification of more effective training strategies for speech-based LLMs.

\section{Conclusion}

This paper presents \textbf{S2SBench}, a benchmark designed to evaluate intelligence degradation in end-to-end Speech Large Language Models. By comparing model performance under audio and text input, S2SBench provides a systematic framework for diagnosing reasoning and generation challenges unique to speech input. We construct diagnostic datasets targeting sentence continuation and commonsense reasoning, and introduce a pairwise evaluation protocol based on perplexity differences to quantify model degradation. Experimental results on Baichuan-Audio demonstrate the benchmark's effectiveness in identifying performance gaps and guiding model improvement.

\section*{Limitations}

This work primarily evaluates the intelligence capability of speech large language models through the speech-to-text (S$\to$T) setting. For Baichuan-Audio, whose output consists of interleaved text and audio tokens, the S$\to$T performance can reasonably reflect the overall end-to-end speech understanding and reasoning ability of the model. However, this evaluation paradigm does not fully capture the model's generative capacity in speech form.

To comprehensively assess the capabilities of speech large language models, a speech-to-speech (S$\to$S) evaluation protocol remains to be further explored. Such an approach would enable direct measurement of both speech comprehension and generation. Nonetheless, the development of S$\to$S evaluation faces technical challenges, particularly due to variations in audio token representations and generation strategies adopted by different models. Establishing standardized benchmarks for S$\to$S evaluation will be an important direction for future research.

\bibliography{custom}

\appendix

% \onecolumn
\section{Visualization of Intelligence Capability During Training}
\label{sec:training-curves}

To gain a deeper understanding of model behavior during training, we visualize the changes in intelligence capability across training steps. These results are shown for both speech-to-text (S$\to$T) and text-to-text (T$\to$T) settings under different training strategies.

\subsection{Single-Stage Training Results}

\begin{figure}[htb]
    \centering
    \begin{subfigure}[b]{0.4\textwidth}
        \includegraphics[width=\textwidth]{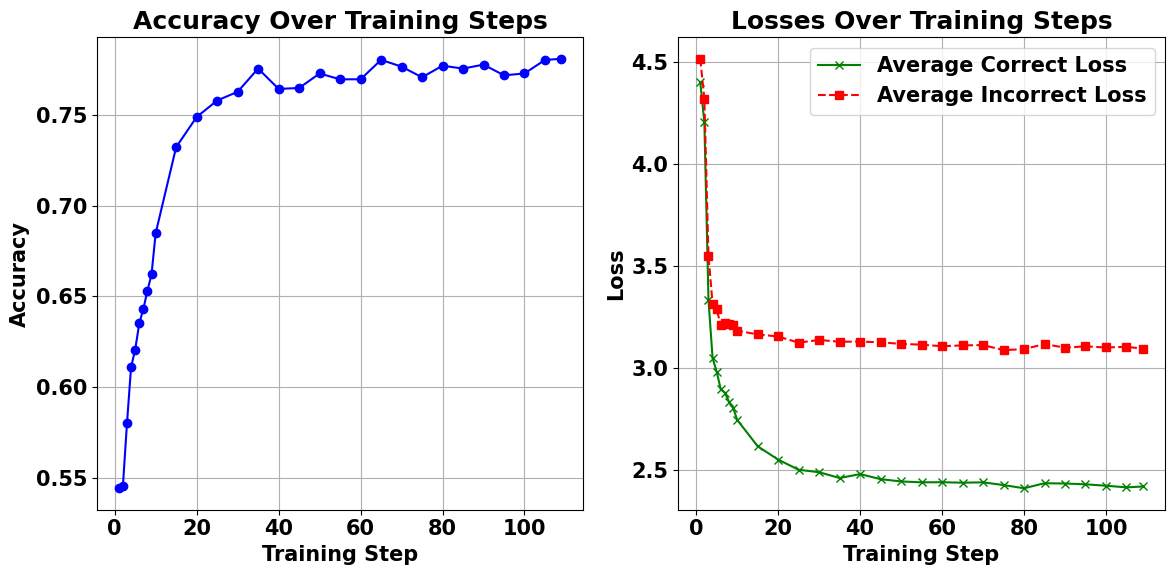} % 添加固定高度
        \caption{sStoryCloze}
    \end{subfigure}
    \hfill
    \begin{subfigure}[b]{0.4\textwidth}
        \includegraphics[width=\textwidth]{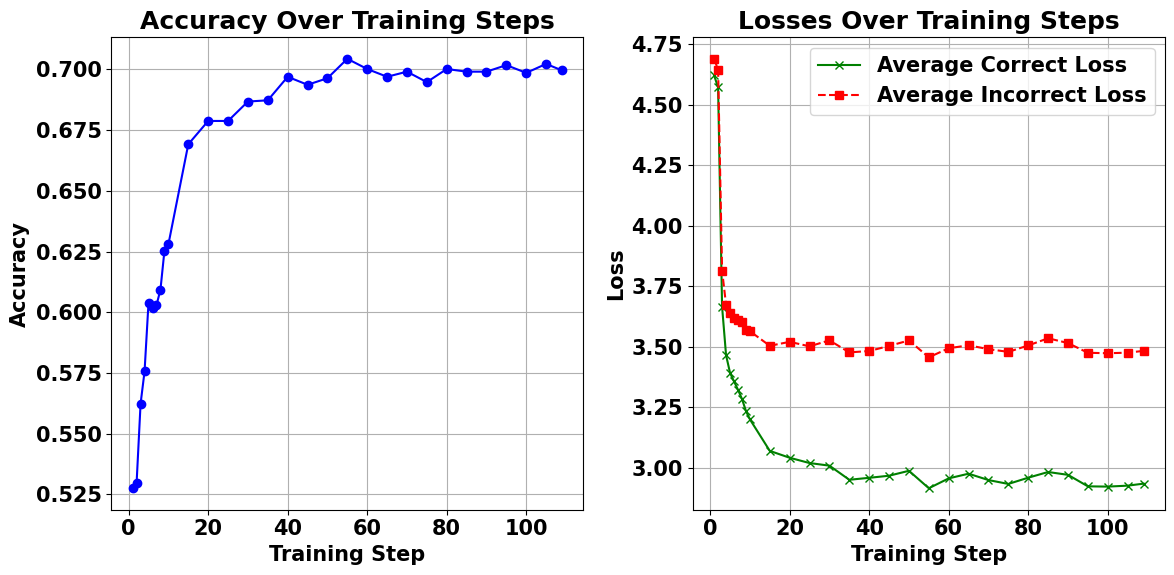}
        \caption{zh-sStoryCloze}
    \end{subfigure}
    
    \begin{subfigure}[b]{0.4\textwidth}
        \includegraphics[width=\textwidth]{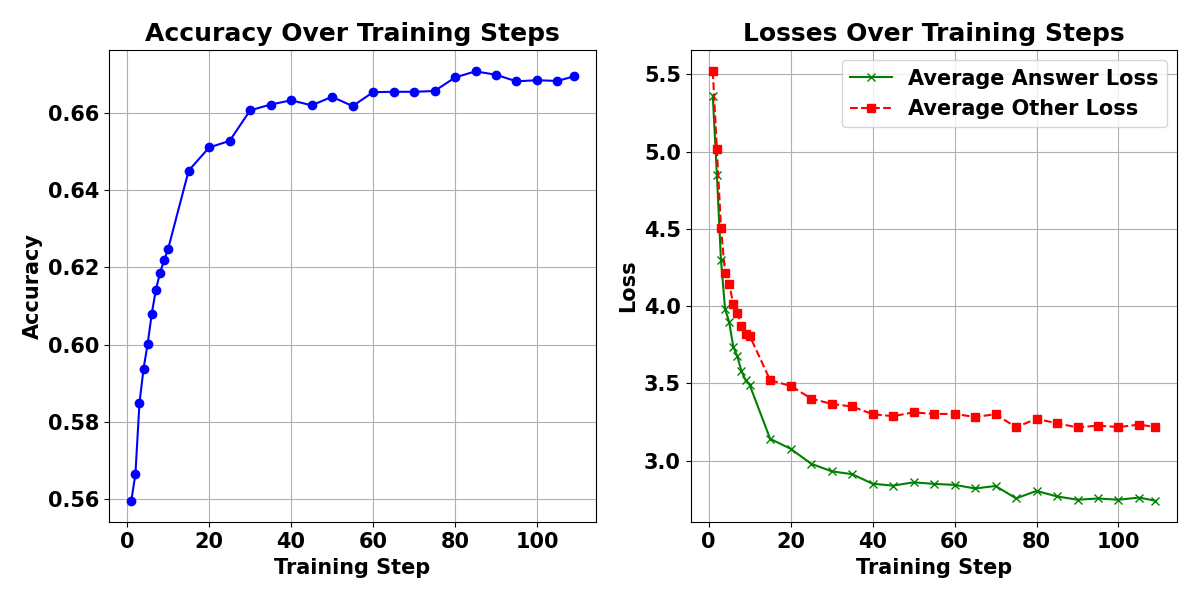}
        \caption{sCMMLU}
    \end{subfigure}
    \caption{Speech-to-text with single-stage training.}
    \label{fig:s2t0}
\end{figure}

\begin{figure}[htb]
    \centering
    \begin{subfigure}[b]{0.4\textwidth}
        \includegraphics[width=\textwidth]{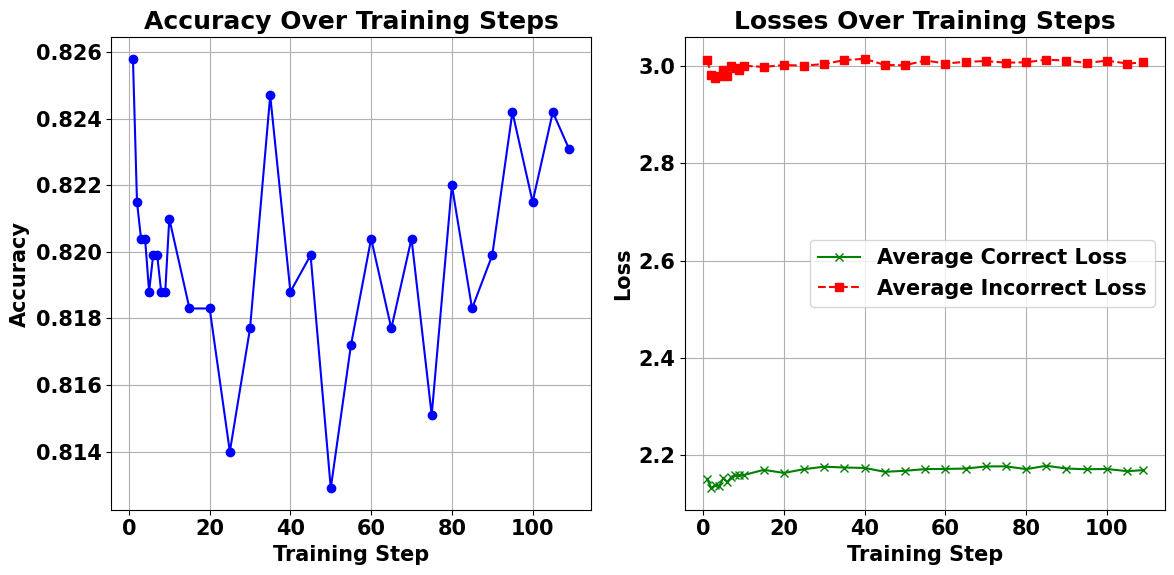}
        \caption{sStoryCloze}
    \end{subfigure}
    \hfill
    \begin{subfigure}[b]{0.4\textwidth}
        \includegraphics[width=\textwidth]{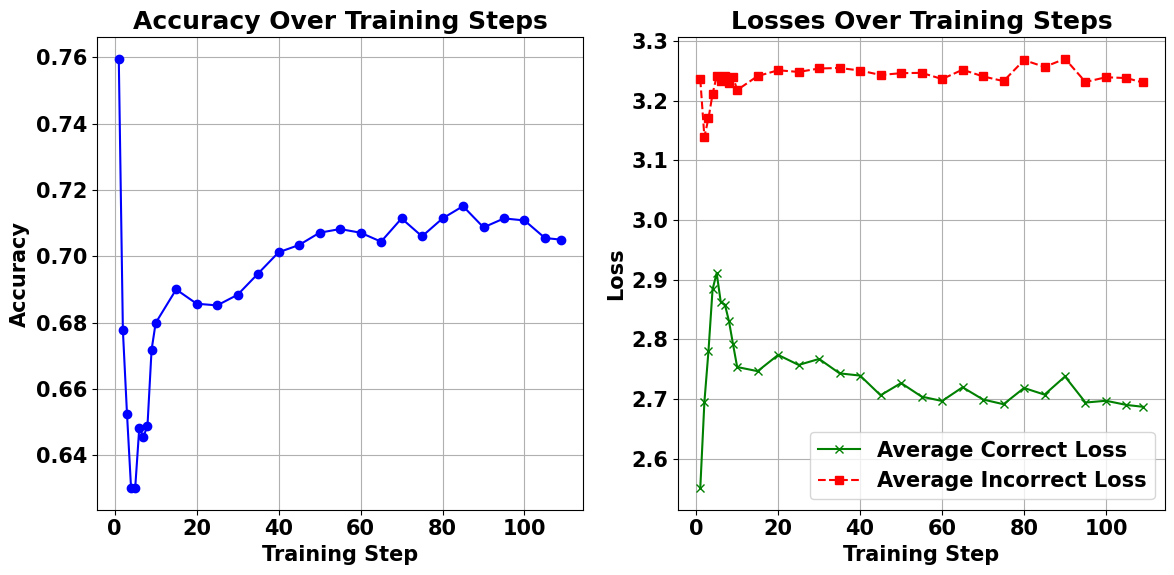}
        \caption{zh-sStoryCloze}
    \end{subfigure}
    \hfill
    \begin{subfigure}[b]{0.4\textwidth}
        \includegraphics[width=\textwidth]{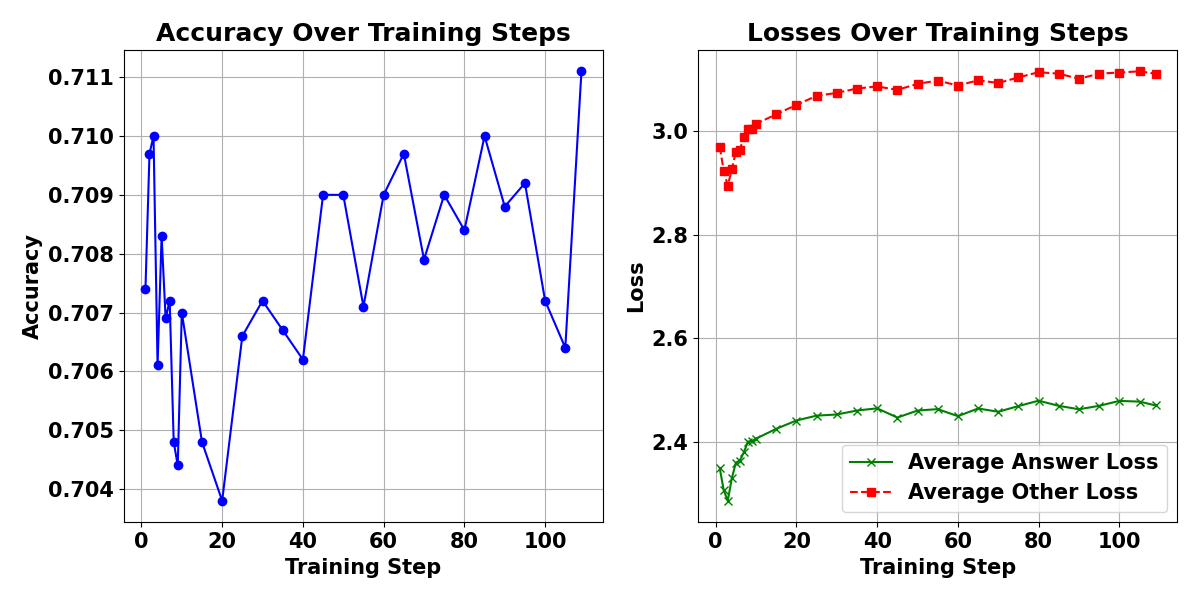}
        \caption{sCMMLU}
    \end{subfigure}
    \caption{Text-to-text with single-stage training.}
    \label{fig:t2t0}
\end{figure}

Figures~\ref{fig:s2t0} and~\ref{fig:t2t0} show the model performance trained with a single-stage strategy in both S$\to$T and T$\to$T modes. In the S$\to$T setting, accuracy improves slowly and loss curves are more volatile. This reflects the difficulty in simultaneously learning from audio tokens while preserving language understanding. The T$\to$T setting yields higher performance and more stable loss curves, confirming text input as the upper bound of reasoning ability.

\subsection{Two-Stage Training Results}

\begin{figure}[H]
    \centering
    \begin{subfigure}[b]{0.4\textwidth}
        \includegraphics[width=\textwidth]{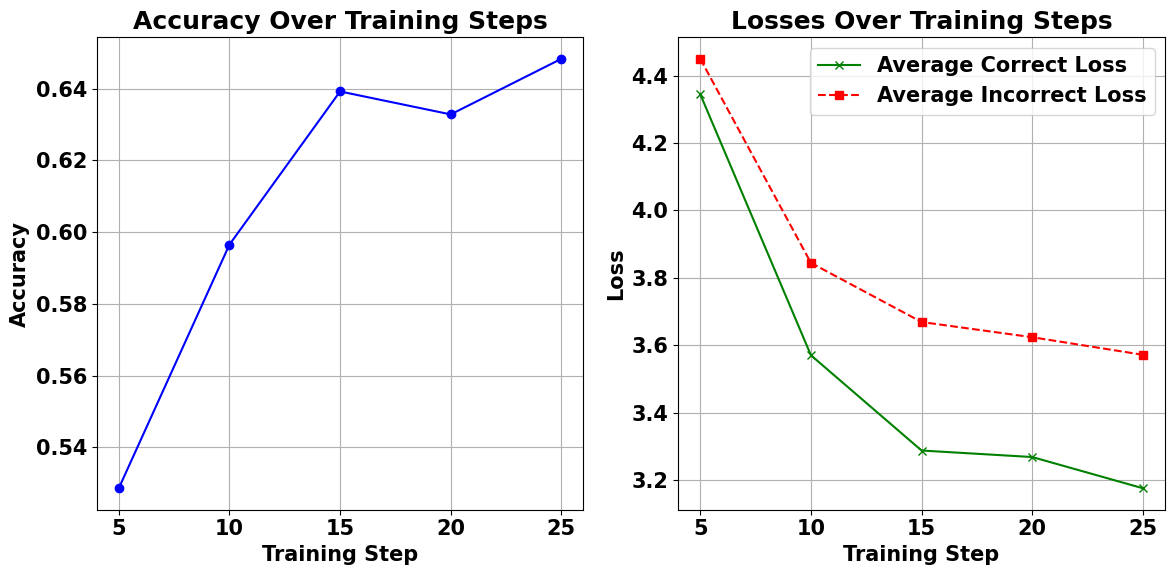}
        \caption{sStoryCloze}
    \end{subfigure}
    \hfill
    \begin{subfigure}[b]{0.4\textwidth}
        \includegraphics[width=\textwidth]{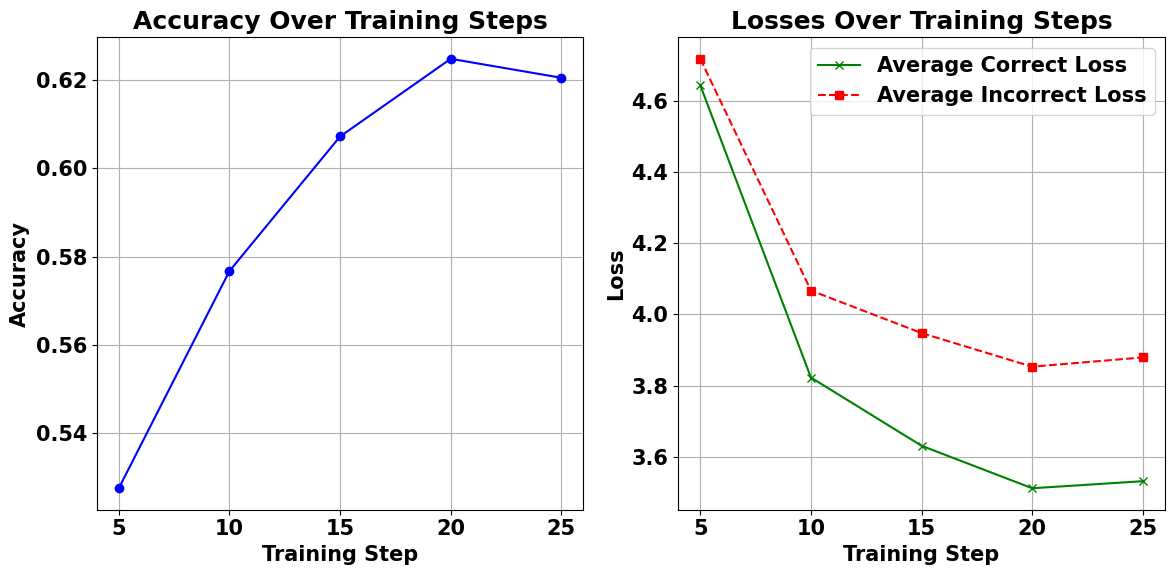}
        \caption{zh-sStoryCloze}
    \end{subfigure}
    \hfill
    \begin{subfigure}[b]{0.4\textwidth}
        \includegraphics[width=\textwidth]{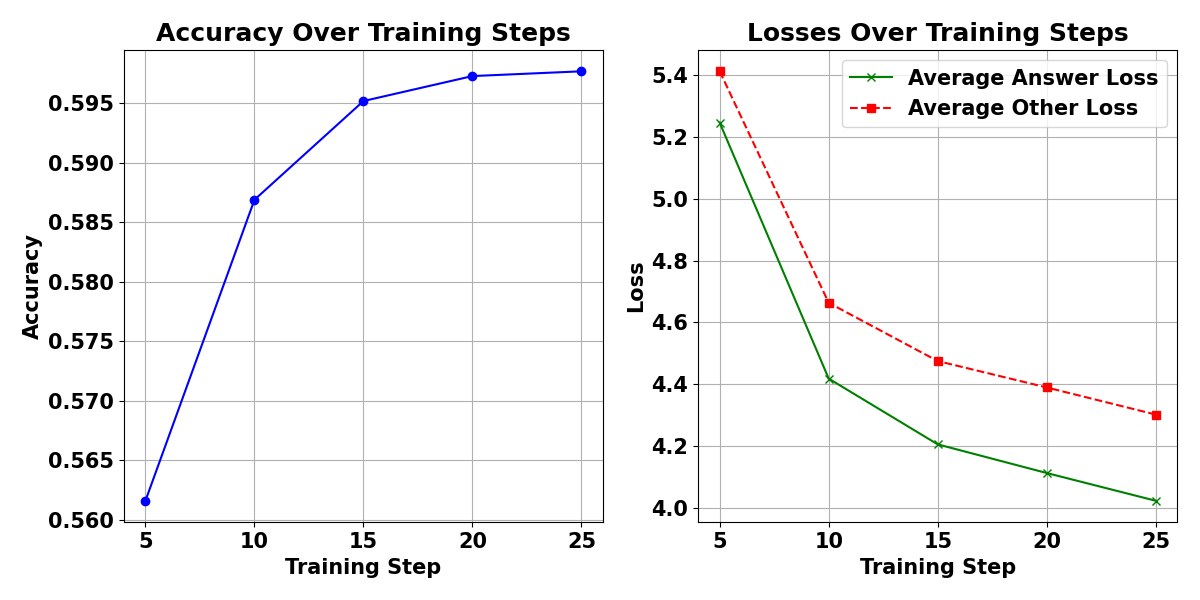}
        \caption{sCMMLU}
    \end{subfigure}
    \caption{Speech-to-text with two-stage training (Stage 1).}
    \label{fig:s2t1}
\end{figure}

\begin{figure}[H]
    \centering
    \begin{subfigure}[b]{0.4\textwidth}
        \includegraphics[width=\textwidth]{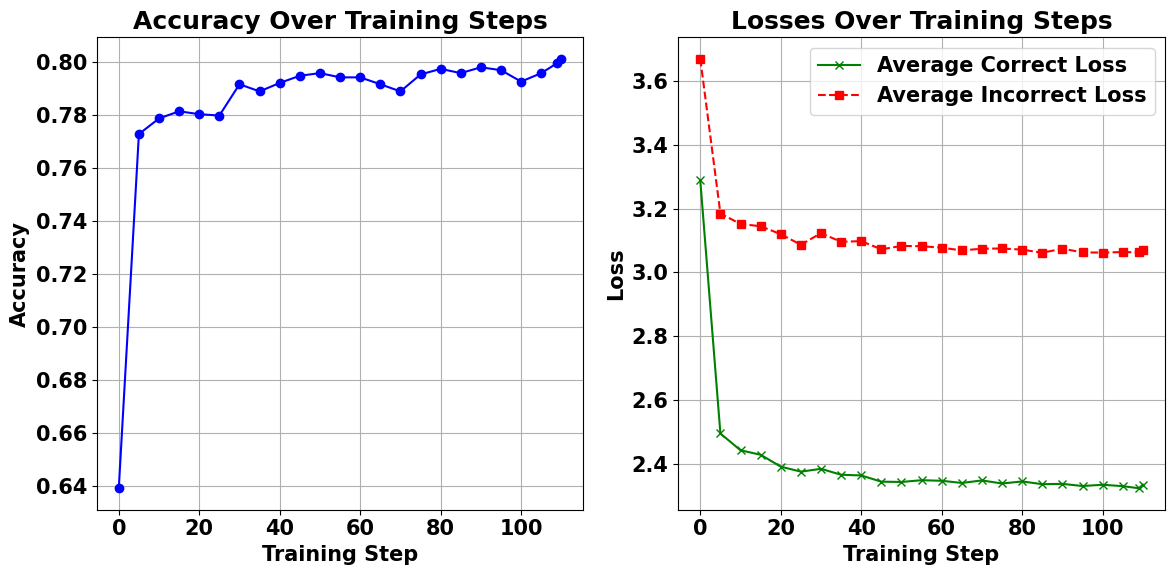}
        \caption{sStoryCloze}
    \end{subfigure}
    \hfill
    \begin{subfigure}[b]{0.4\textwidth}
        \includegraphics[width=\textwidth]{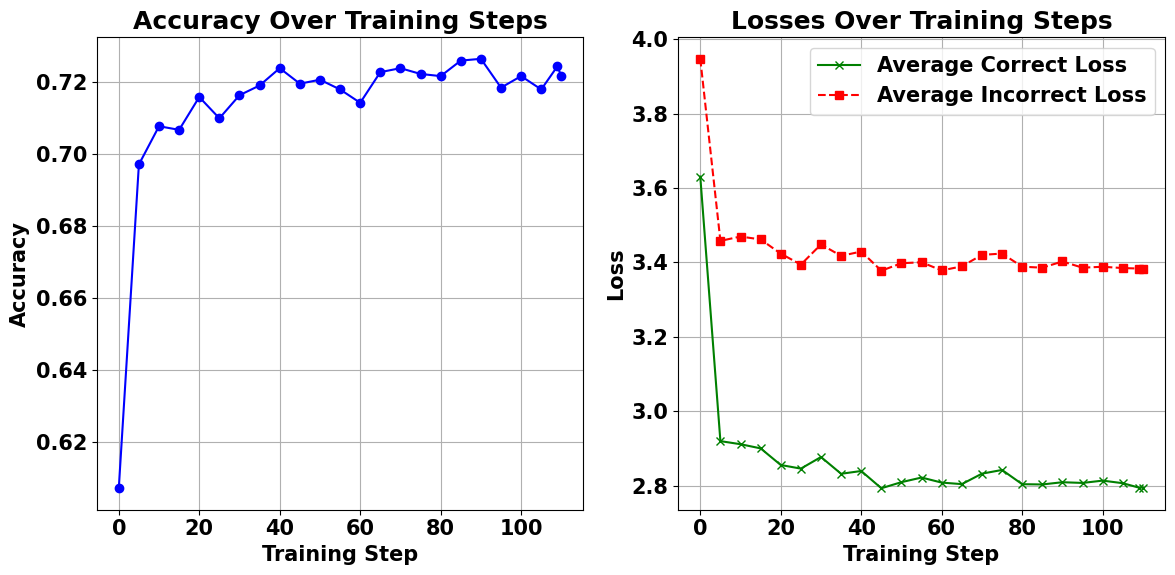}
        \caption{zh-sStoryCloze}
    \end{subfigure}
    \hfill
    \begin{subfigure}[b]{0.4\textwidth}
        \includegraphics[width=\textwidth]{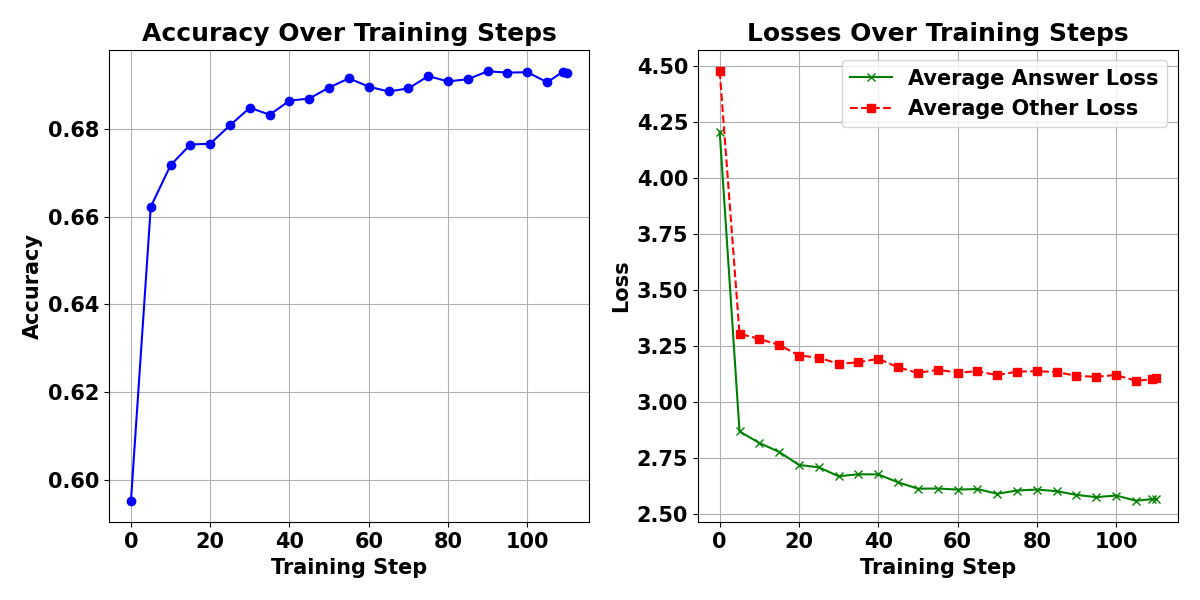}
        \caption{sCMMLU}
    \end{subfigure}
    \caption{Speech-to-text with two-stage training (Stage 2).}
    \label{fig:s2t2}
\end{figure}

\begin{figure}[H]
    \centering
    \begin{subfigure}[b]{0.4\textwidth}
        \includegraphics[width=\textwidth]{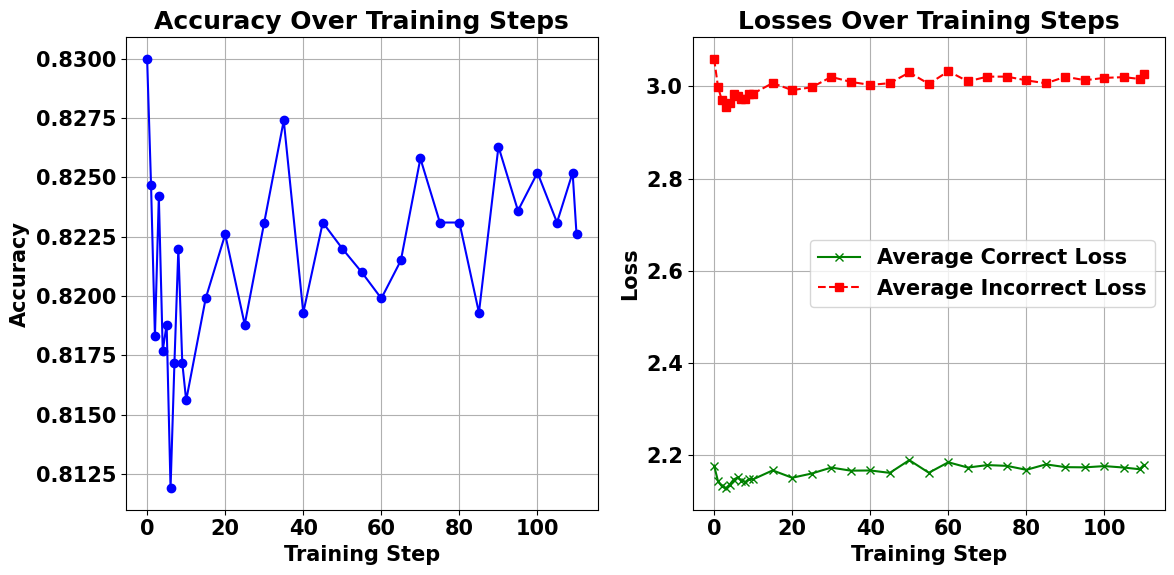}
        \caption{sStoryCloze}
    \end{subfigure}
    \hfill
    \begin{subfigure}[b]{0.4\textwidth}
        \includegraphics[width=\textwidth]{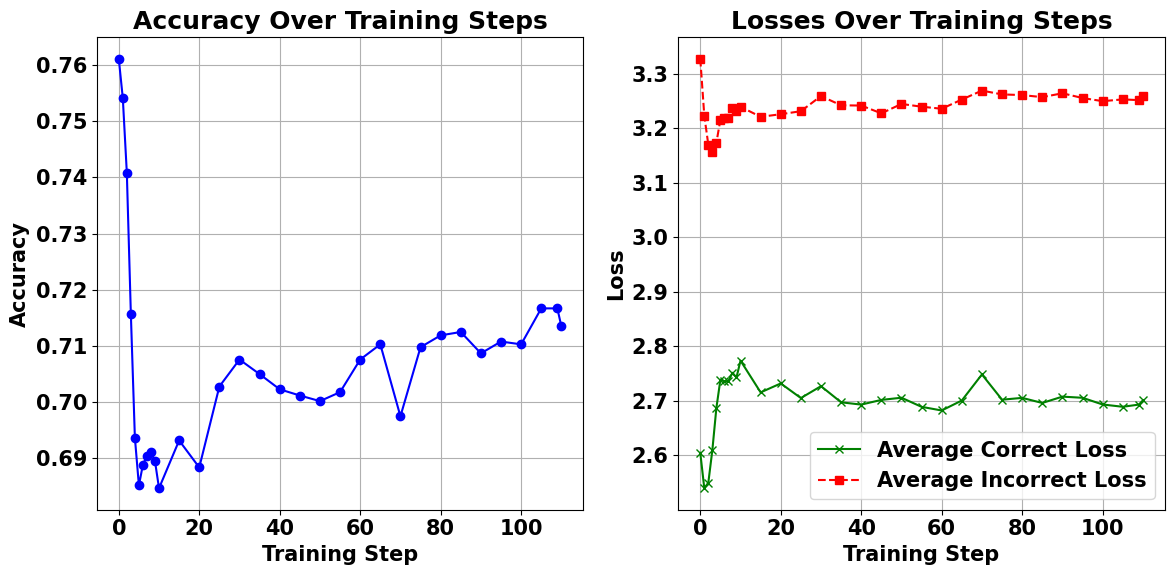}
        \caption{zh-sStoryCloze}
    \end{subfigure}
    \hfill
    \begin{subfigure}[b]{0.4\textwidth}
        \includegraphics[width=\textwidth]{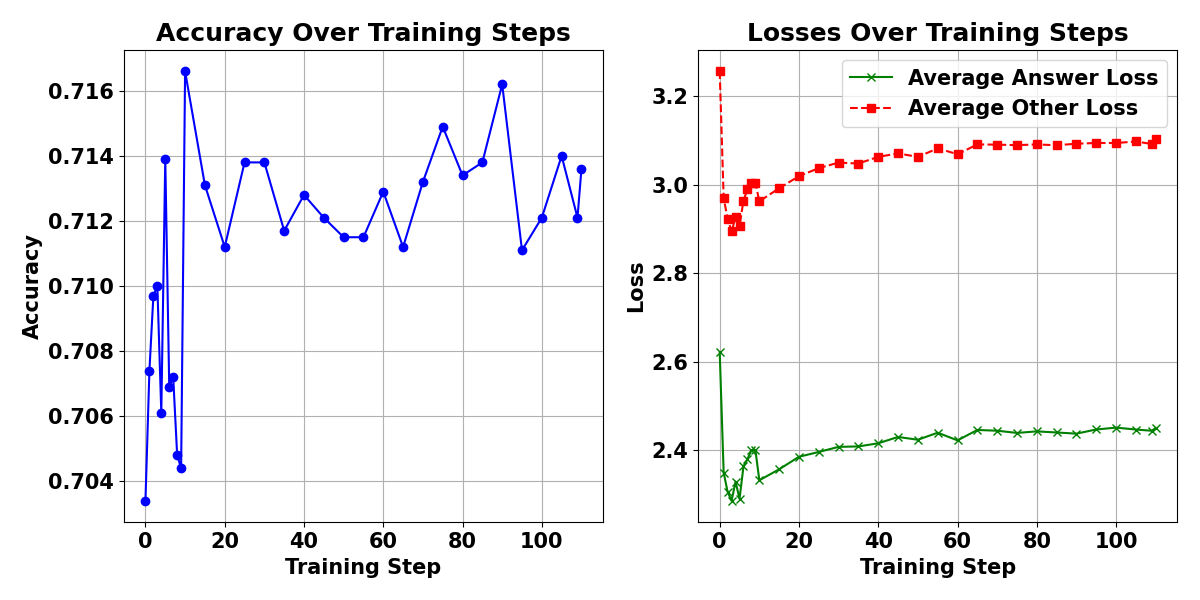}
        \caption{sCMMLU}
    \end{subfigure}
    \caption{Text-to-text with two-stage training (Stage 2).}
    \label{fig:t2t2}
\end{figure}

Figures~\ref{fig:s2t1}, \ref{fig:s2t2}, and~\ref{fig:t2t2} display performance under the two-stage training strategy. Compared to single-stage training, models exhibit significantly better performance in S$\to$T mode. The gap between positive and negative loss becomes more distinguishable, indicating stronger reasoning ability. Accuracy increases more rapidly and consistently, especially on the sStoryCloze dataset.

\subsection{Analysis Summary}

In the T$\to$T setting, performance on the sCMMLU dataset shows noticeable improvement across training. This can be attributed to the presence of commonsense-related text data in the pretraining corpus, which allows the language model to better handle such knowledge-intensive tasks during fine-tuning. On the other hand, we observe that the intelligence degradation on the zh-sStoryCloze task is more severe than on sStoryCloze, even in the T$\to$T setting. This may be due to weaker pretrained semantic alignment in Chinese data, or greater challenges in modeling discourse coherence in the Chinese language under limited resources.

In summary, the visualizations demonstrate that the two-stage training strategy significantly enhances model stability and performance, especially in speech-based tasks. Speech-to-text models trained with this strategy not only achieve higher accuracy but also maintain clearer separation between positive and negative samples. Furthermore, commonsense reasoning ability benefits from the pretrained knowledge encoded in large-scale text corpora, while discourse-level understanding in Chinese remains more challenging. The proposed evaluation framework effectively captures these trends and provides fine-grained insight into the training dynamics of speech-enabled large language models.

\end{document}